\documentclass[12pt]{iopart}
\usepackage{iopams}
\usepackage{graphicx}
\DeclareGraphicsExtensions{.pdf,.png,.jpg}
\usepackage{bm}
\usepackage{dcolumn}
\usepackage{psfrag}
\usepackage{braket}
\usepackage{url}
\def\be {\begin{equation}}
\def\ee {\end{equation}}
\def\bea {\begin{eqnarray}}
\def\eea {\end{eqnarray}}
\def\bc {\begin{center}}
\def\ec {\end{center}}
\def\bfg {\begin{figure}}
\def\efg {\end{figure}}
\def\bi {\begin{itemize}}
\def\ei {\end{itemize}}
\def\nnr {\nonumber\\ }
\def\nn {\nonumber}

\def\vs {\vspace}

\def\Re {{\rm Re}}
\def\Im {{\rm Im}}
\def\Z{{\mathbb Z}} \def\R{{\mathbb R}}  
\def\C{{\mathbb C}}   

\begin{document}
\voffset=-0.5cm
\hoffset=+0.5cm
\title[GUP Corrections to the Simple Harmonic Oscillator in Phase Space]{Generalized Uncertainty Principle Corrections to the Simple Harmonic Oscillator in
Phase Space}

\author{Saurya Das,\ Matthew P.G. Robbins,\ Mark A. Walton\footnote{Corresponding author.}}

\address{Department of Physics and Astronomy, University of Lethbridge, Lethbridge, Alberta, T1K 3M4, Canada }

\eads{ \mailto{saurya.das@uleth.ca}, \mailto{matthew.robbins@uleth.ca}, \mailto{walton@uleth.ca} }

\begin{abstract}
We compute Wigner functions for the harmonic oscillator including corrections from
generalized uncertainty principles (GUPs), and
study the corresponding marginal probability densities and other properties.
We show that the GUP corrections to the Wigner functions can be significant, and comment on their potential
measurability in the laboratory.
\end{abstract}

\date{\today}

\section{Introduction}

It is currently not possible to access the natural energy scale of quantum gravity, the Planck energy.  It is feasible, however to consider  low-energy effects, e.g., the quantum-gravity induced perturbative corrections to non-relativistic quantum mechanics.
One avenue is the study of corrections to the Schr\"odinger equation originating from the GUP proposed in
various candidate theories of quantum gravity (such as string theory, loop quantum gravity,  etc.).  A modification is postulated of the usual Heisenberg algebra (and the resulting  Heisenberg uncertainty principle),
to\footnote{Here and throughout, $\hat a$ denotes an operator observable, and $a$ the corresponding c-number.}
\be
[\hat x_i, \hat p_j]\  =\  i\hbar\, f_{i,j}(\hat p)\ .
\label{GUPfp}
\ee
For the 1-dimensional case considered in this paper,  $f_{i,j}$ becomes a single function, $f(p)$.  In \cite{kmm},  the quadratic form $f( p)=1+\beta  p^2$ was suggested, while in \cite{adv},
a linear $+$ quadratic function, \be f( p)\  =\ 1\,+\,\alpha p \,+\, \beta  p^2\,, \  \ \label{fpab}\ee  was proposed.
Here $\alpha=\alpha_0/{M_{\rm Pl}\,c} = \alpha_0\ell_{\rm Pl}/\hbar, M_{\rm Pl}=$ Planck mass,
$\ell_{\rm Pl} \approx 10^{-35}~$metre\,= Planck length. $\alpha_0$ can be assumed to be order unity,
and $\beta = {\cal O} (\alpha^2)$.
%
%
%

Over the years, various modifications of the canonical commutation relations have been considered, with many different  motivations.\footnote{Motivations include the so-called Wigner problem \cite{WigProb}, the related Feynman problem \cite{FeynProb}, and quantum groups, for examples.} We focus on (\ref{GUPfp}, \ref{fpab}) because we are ultimately interested in the low-energy effects of quantum gravity, and because, in that context, modifications (\ref{GUPfp}, \ref{fpab}) are quite general.
The form (\ref{fpab}) of $f(p)$ has been suggested by various approaches to quantum gravity, as well as from black hole physics and doubly special relativity theories \cite{qgtheories}.
Various perturbative and non-perturbative effects of the correction terms were
studied in a number of papers including  those for low energy systems,  {the} fundamental nature of spacetime,  {and cosmology} (for a related review, see \cite{hoss}; see also references therein).

Naturally, one of the first examples studied in this context was the harmonic oscillator,
in which GUP corrections to the eigenvalues and eigenfunctions were computed \cite{kmm,adv}.\footnote{Recently, the methods of supersymmetric quantum mechanics have also been applied to the GUP-modified harmonic oscillator \cite{APN}.}
It is  {anticipated} that effects of at least some of these corrections may be observable in the
low energy laboratory, for example in quantum optics.

To explore this further, in this paper we study the GUP corrections to the harmonic oscillator in
phase space, and in particular compute and plot the Wigner functions corresponding to the
unperturbed and perturbed eigenfunctions for various $n$, and  {then} study their differences. We note that,
depending on the value of $\alpha_0$, these differences could be significant, and therefore in
principle may have observational consequences. In the following sections, we briefly review
Wigner functions, and compute and plot them for the problem described above. In the concluding section,
we comment on potential applications.

\section{Wigner Functions}
\label{Wigner Function}

Rather than using the operator formalism, it is possible to work with a phase-space formulation of quantum mechanics, developed by Groenewold and Moyal.  In it, observables are represented by (generalized) functions  {in} phase space, that are multiplied using an associative (Moyal) star product,
\bea
\star\ =\  \exp\left\{\frac{i\hbar}{2}\left(\overleftarrow{\partial_x}\,\overrightarrow{\partial_p}
-\overleftarrow{\partial_p}\,\overrightarrow{\partial_x}\right)\right\}\ , \label{MoyalStar}
\eea
and states are described by the  {well-known} Wigner function (see \cite{psqm}, e.g., for recent reviews, and \cite{pedpsqm} for pedagogical treatments).  The Wigner transform maps an operator $\hat g$ to the corresponding phase-space function,
\be
{\cal{W}}(\hat g)=\hbar\int dy\, e^{-ipy}\left\langle x+ {\hbar y}/{2}\big|\, \hat g \,\big|x- {\hbar y}/{2}\right\rangle\ ,
\ee
such that the star product of observables in phase space is homomorphic to the operator product,
\be
{\cal W}\big( \hat a\, \hat b \big)\ =\ {\cal W}(\hat a)\, \star\, {\cal W}(\hat b)\ .
\label{homo}
\ee
Up to a multiplicative constant, the Wigner function is nothing but the Wigner transform of the density matrix $\hat\rho$:
\be	\fl W(x,p)\, =\,  \frac{{\cal{W}}(\hat{\rho})}{2\pi\hbar} \,=\, \frac{{\cal{W}}(\ket{\psi}\bra{\psi})}{2\pi\hbar} \,=\, \frac{1}{2\pi}\int_{-\infty}^{\infty}\,\psi^{*}\left(x- {\hbar y}/{2}\right)\,\psi\left(x+ {\hbar y}/{2}\right)\,e^{-i py}\,dy\ .
	\label{WF - x}
\ee
Here $\hat{\rho}$ is the density matrix, $\psi$ is the wave function in $x$-space, $x$ is the position, and $p$ is the momentum. The Wigner function can also be found using the wave function, $\phi$, in $p$-space:
\bea
	W(x,p)=\frac{1}{\hbar}\int_{-\infty}^{\infty}\phi^{*}\left(p+ {u}/{2}\right)\,\phi\left(p- {u}/{2}\right)\,e^{i xu/\hbar}\,du\ .
	\label{WF - p}
\eea

\begin{center}
\begin{figure}
\includegraphics[scale=0.26,angle=0]{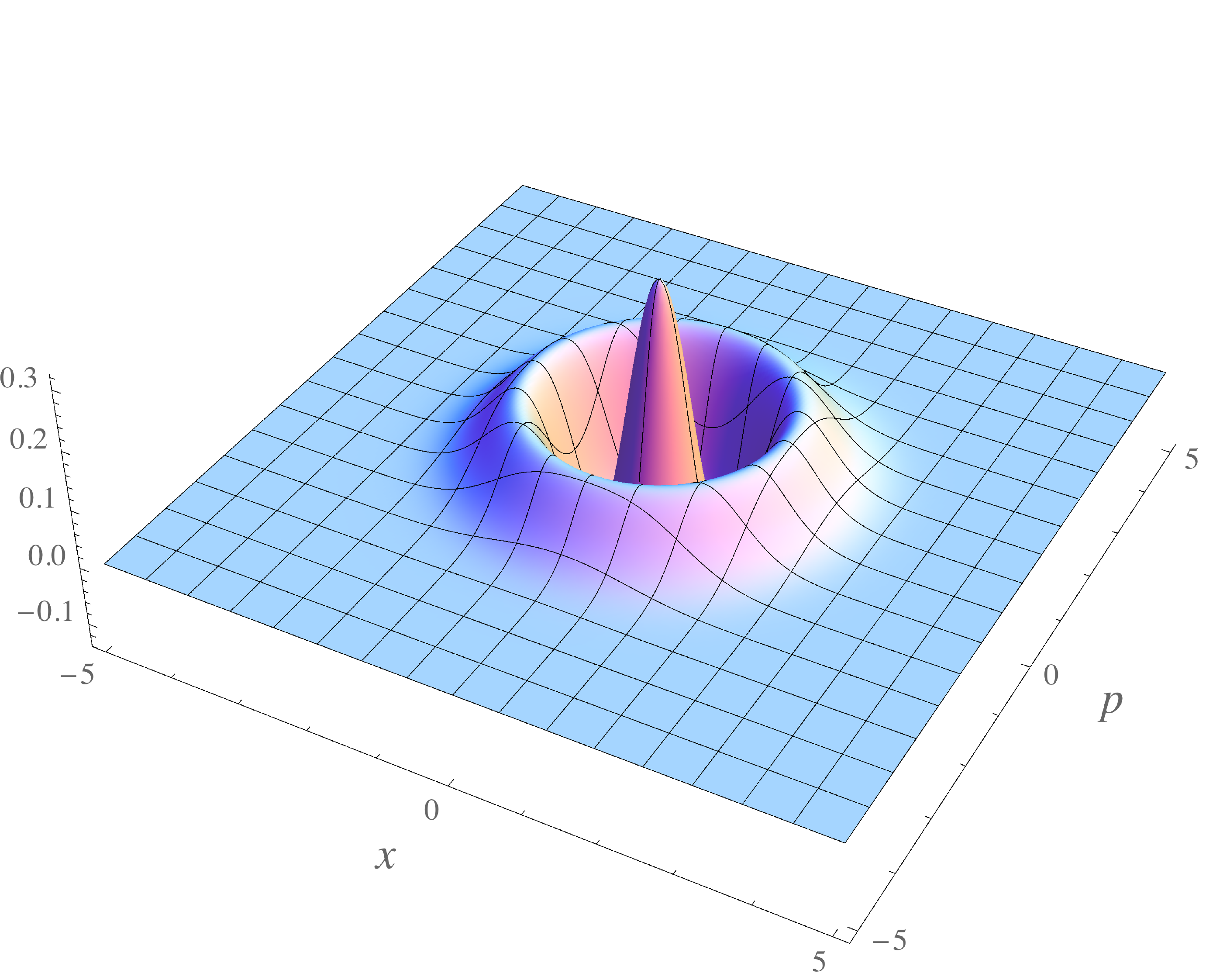}\qquad
\includegraphics[scale=0.26,angle=0]{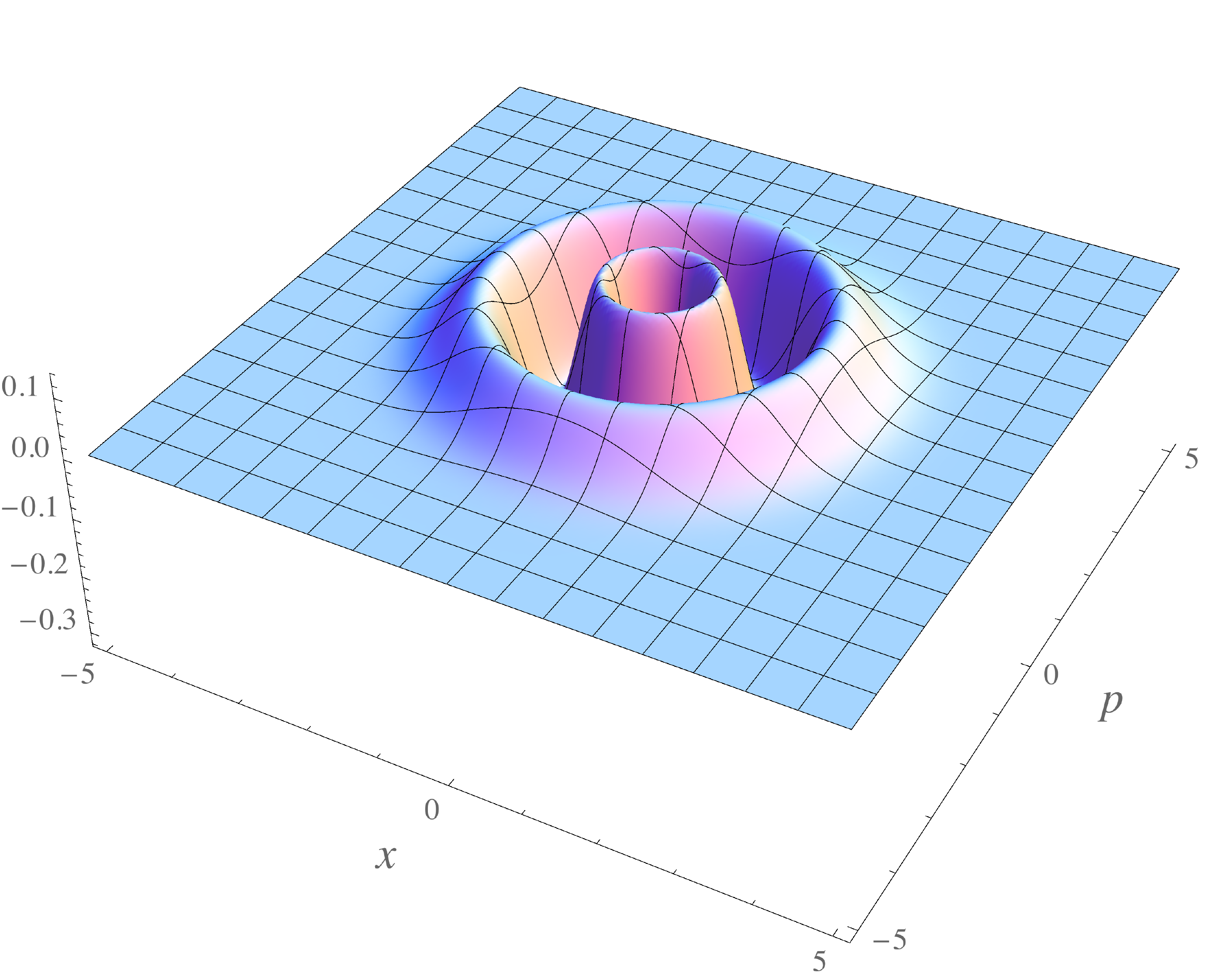}
\caption{The simple harmonic oscillator Wigner functions for $n=2$ (left) and $n=3$ (right).  With $m=\hbar=\omega=1$, they  only depend on $x^2+p^2$; the circular symmetry is evident. Because the Wigner function can be negative, it is known as a quasi-probability distribution.}
\label{SHO WF}
\end{figure}
\end{center}

 {One other} alternative method to find the Wigner function is to solve the stargenvalue equations
\bea
H\star W(x,p)&\ =\ E\,W(x,p)\ ,\\
W(x,p)\star H&\ =\ E\,W(x,p)\  \ .
\label{stargenvalue}
\eea
$H$ is the Hamiltonian of the system, and $E$ is the energy.\\
\indent For example, Figure \ref{SHO WF} displays the Wigner functions of two energy
eigenstates of the simple harmonic oscillator.\\
\indent Important properties of the Wigner function include:
(i) reality, $W(x,p)=W(x,p)^{*}$,
(ii) position probability density, $P(x)=|\psi(x)|^2=\int W(x,p)\,dp$,
(iii) momentum probability density, $P(p)=|\phi(p)|^2=\int W(x,p)\,dx$, and
(iv) normalization, $\int W(x,p)\,dx\,dp=1$. Using the Wigner function, the expectation value $\langle a \rangle$ of an operator $\hat a$ is
\be
\langle  a \rangle\ =\ \int\, W(x,p)\, a(x,p)\,\,dx\,dp\ ,
\ee
where $a(x,p) = {\cal W}(\hat a)$  is the Weyl transform of $\hat a$.

The equivalence of phase space quantum mechanics to the operator formulation follows from the Wigner transform ${\cal W}$, and its inverse, ${\cal W}^{-1}$, known as the Weyl map. The latter's relation to Weyl operator ordering is made plain by expanding  \be {\cal W}^{-1}\left(  e^{\zeta x + \eta p}  \right) \ =\   e^{\zeta \hat x + \eta \hat p}\ \ee in powers of $\zeta$ and $\eta$. This last equation also indicates how general functions  {in} phase space map to operators: Fourier component by component.

Using $[\hat x, \hat p]=i\hbar$ and a simple Baker-Campbell-Hausdorff formula, one finds
\be
\fl \exp\left(  \zeta\hat x + \eta\hat p \right)\, \exp\left(  \zeta'\hat x + \eta'\hat p \right)\ =\ \exp\left[  (\zeta+\zeta')\hat x + (\eta+\eta')\hat p \right]\, \exp\left[ i\hbar (\zeta\eta' -\eta\zeta')/2\right]\ ,
 \label{HWgroup}
 \ee
 the defining relation of the Heisenberg-Weyl group.  Then (\ref{homo}) leads to the form (\ref{MoyalStar}) of
the Moyal star product.

If the Heisenberg commutation relations are generalized to $[\hat x, \hat p]=i\hbar(1 + \alpha \hat p + \beta\hat p^2)$, then a similar computation yields a modified GUP star product
\be
\log\left(\, \tilde \star\,\right)\ =\  \log\left(\, \star\,\right)\, \cdot\,\left\{\, F_0 \,+\, \frac{F_1}{6}\, \left( \overleftarrow{\partial_x} - \overrightarrow{\partial_x}\right)\, -\, \frac{F_2}{12}\, \overleftarrow{\partial_x}\,\overrightarrow{\partial_x}\, +\, \ldots \,\right\}\  . \label{GUPstar}
\ee
Here
\be
F_n\ :=\ \left[ i\hbar f(p) \frac{d}{dp} \right]^n\, f(p)\ ,
\ee
and the exponent in (\ref{GUPstar}) does not terminate for polynomial $f(p)$, such as (\ref{fpab}). This GUP star product  encodes completely the effects of the GUP in phase-space quantum mechanics. As a simple example, the $\tilde\star$-commutator realizes the generalized commutation relation $ x\tilde\star p - p\tilde\star x = i\hbar f(p)$.

Clearly, it is impracticable to solve equations (8)-(\ref{stargenvalue}) for the GUP-corrected Wigner functions, if the GUP star product $\tilde\star$  is used.   We will instead take the simpler approach of  finding the GUP-corrected wave functions in momentum space first  (building on the work of \cite{kmm}), and then use (\ref{WF - p}) to calculate the Wigner functions.

\section{Corrections to harmonic oscillator from quadratic GUP}
\label{One-parameter GUP corrections}

We will first review the work of \cite{kmm}
in which $f(p)=1+\beta p^2$, the simplest case of the previously mentioned quantum gravity phenomenologies ($\alpha=0$), for which the GUP assumes the form
\bea
\Delta x\Delta p
\geq
\frac{\hbar}{2}\left(1 + \beta\langle{p^2}\rangle\right)\,\ .
\label{KMM UP}
\eea
For small $\beta$, with
$\displaystyle{\epsilon=2E/m\hbar^2\omega^2}$
and $\displaystyle{\eta^2=1/(m\hbar\omega)^2}$\,,
the GUP-corrected Schr\"odinger equation for the harmonic oscillator in momentum space becomes
\cite{kmm}
\bea
	\frac{d^2\phi(p)}{dp^2}+\frac{2\beta p}{1+\beta p^2}\frac{d\phi(p)}{dp}+\frac{1}{\left(1+\beta p^2\right)^2}\left[\epsilon-\eta^2p^2\right]\phi(p)=0\,.
	\label{KMM DE}
\eea
The solution is \cite{kmm}
\bea
\phi(p) \,=\,  \frac{{\cal N}}{\left(1+\beta p^2\right)^{\sqrt{q+r}}}\ {}_2F_1\left(a,b;c;\frac{1}{2}+i\frac{\sqrt{\beta}}{2}p\right)\,,
	\label{KMM solution}
\eea
where
$
q=\epsilon/4\beta,\
r=\eta^2/4\beta^2,\
a=\frac{1}{2}\left(1-\sqrt{1+16r}\right)-2\sqrt{q+r},\
b=\frac{1}{2}\left(1+\sqrt{1+16r}\right)-2\sqrt{q+r},\
c=1-2\sqrt{q+r}.
$
The solution is normalizable, with normalization constant ${\cal N}$, if $b=-n\in\Z^{-}\cup\{0\}$ ($a$, $c$, and $\sqrt{q+r}$ can each be expressed in terms of $n$). The energy eigenvalues are then \cite{kmm}
\bea
E_n=\hbar\omega\left(n+\frac{1}{2}\right)
\left(\frac{\beta}{2\eta}+\sqrt{1+\frac{\beta^2}{4\eta^2}}\right)+\hbar\omega\frac{\beta}{2\eta}n^2\ .
\label{EnB}\eea

\begin{center}
\begin{figure}
\includegraphics[scale=0.26,angle=0]{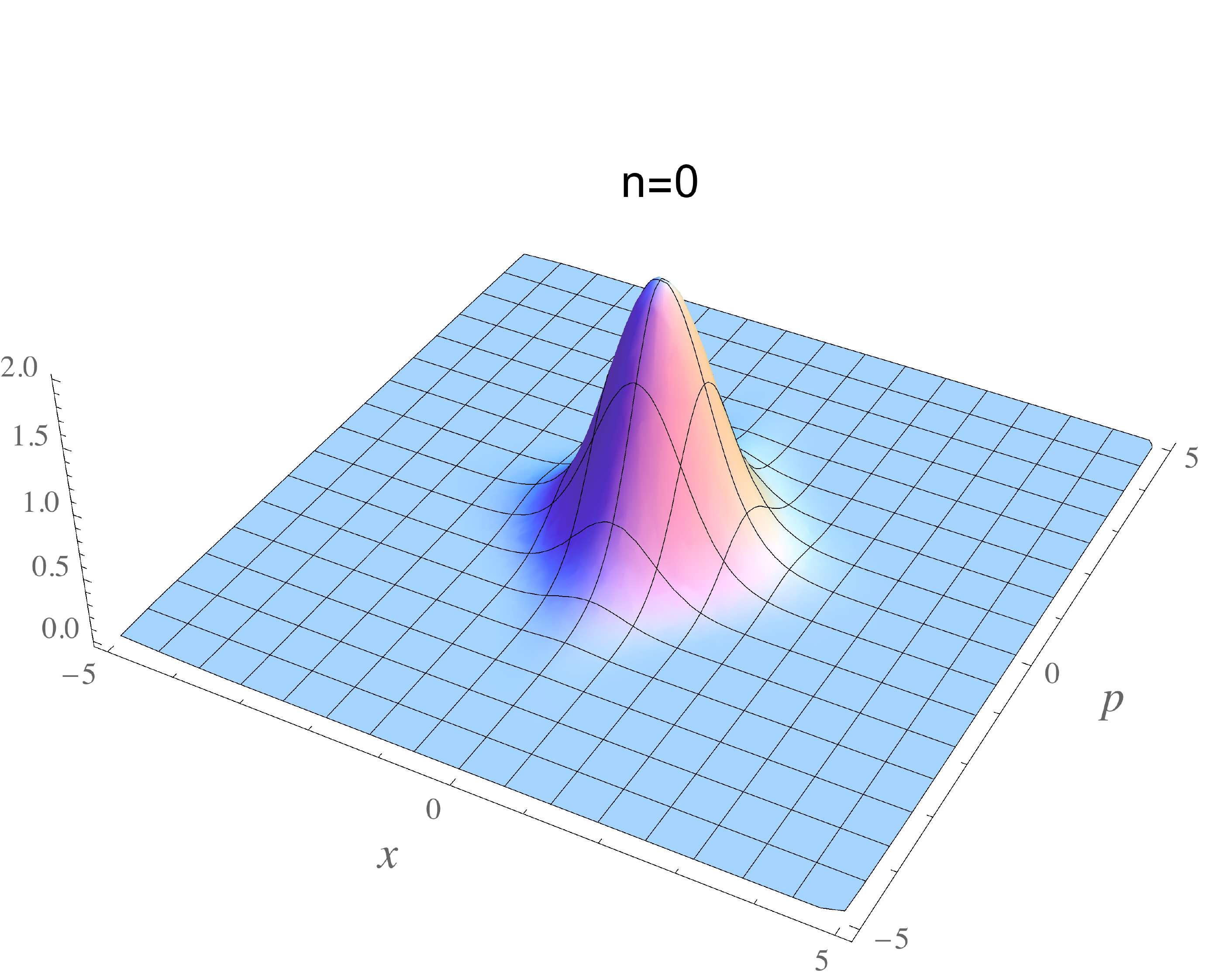}\qquad
\includegraphics[scale=0.26,angle=0]{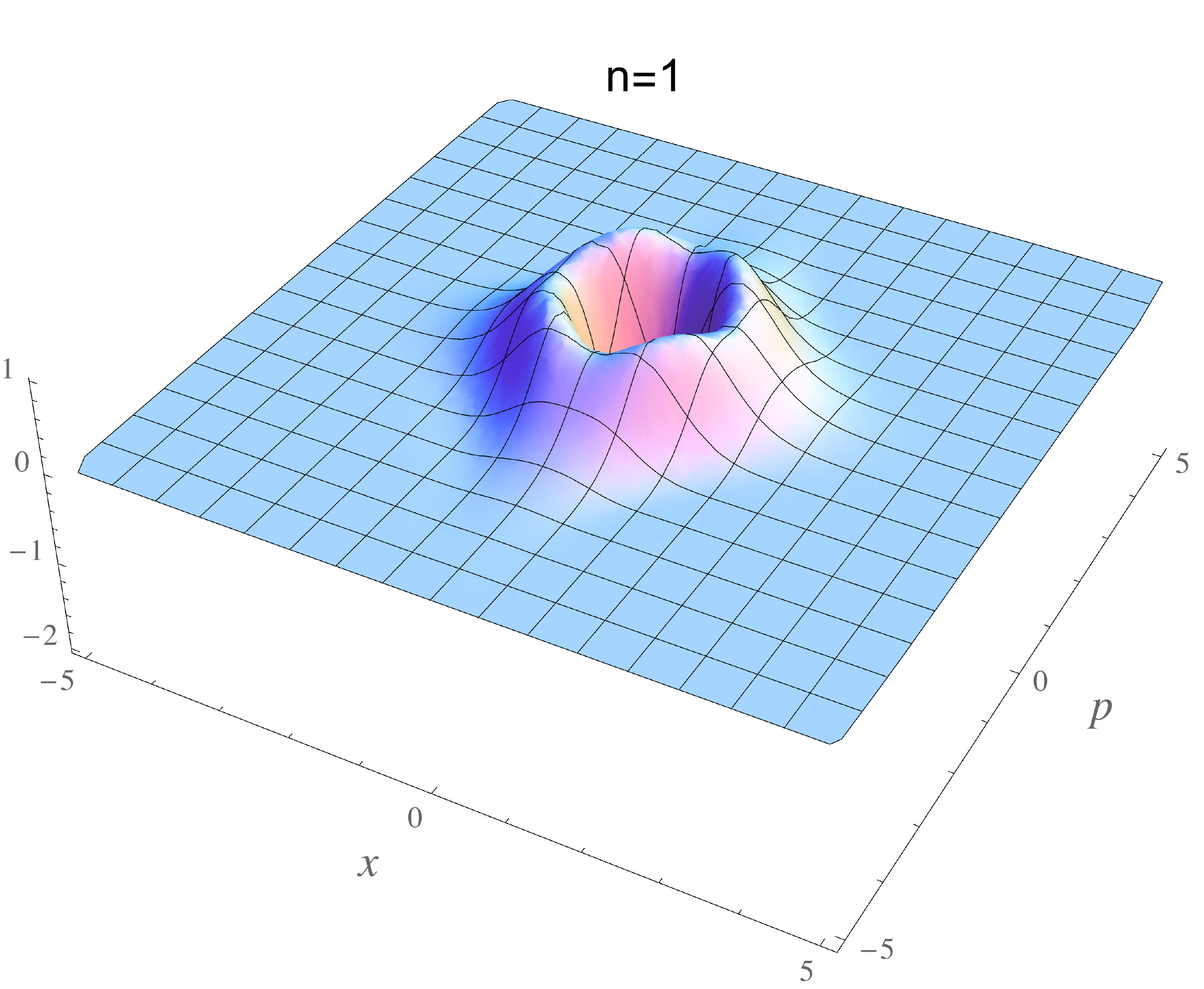}\\
\includegraphics[scale=0.26,angle=0]{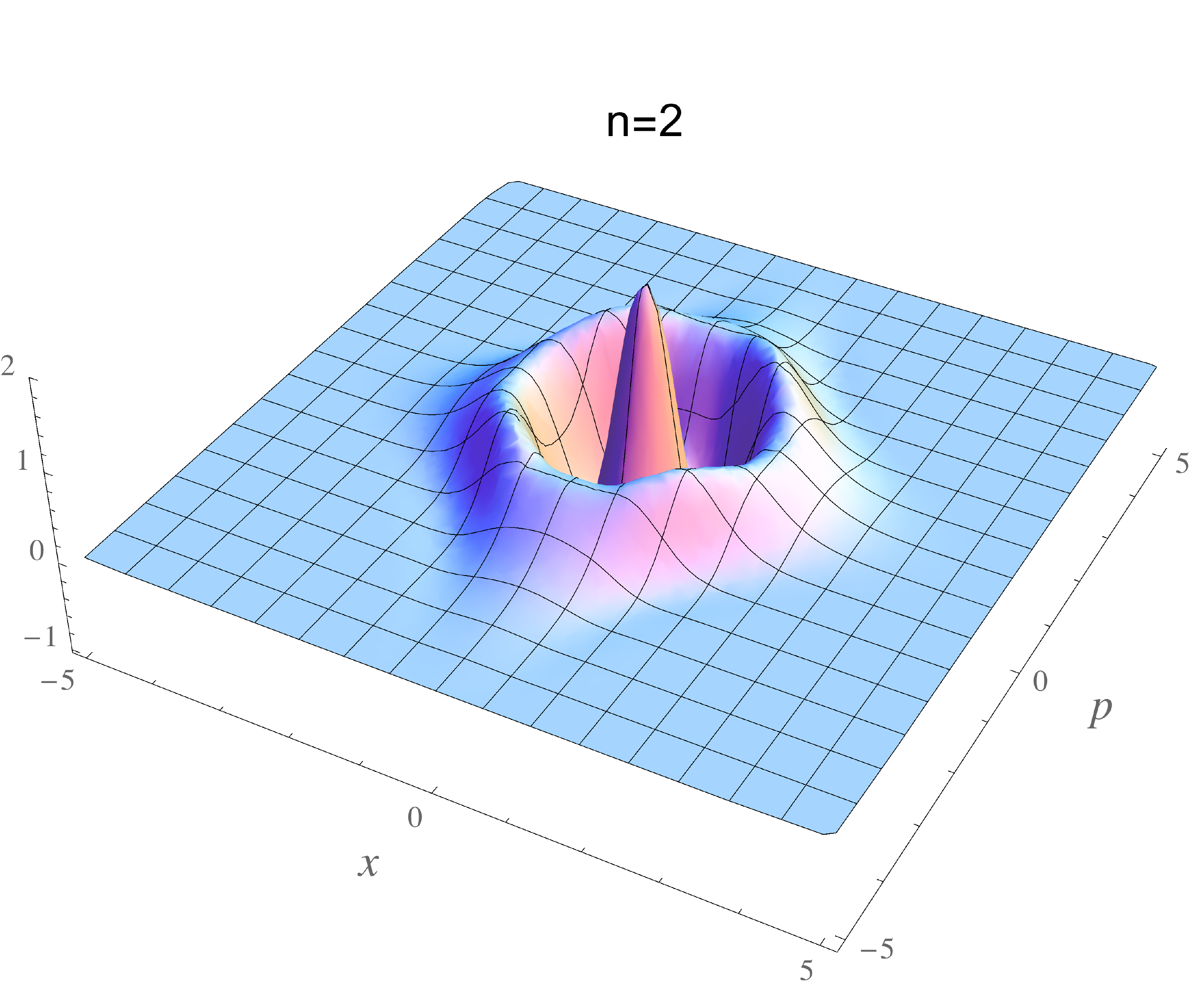}\qquad
\includegraphics[scale=0.26,angle=0]{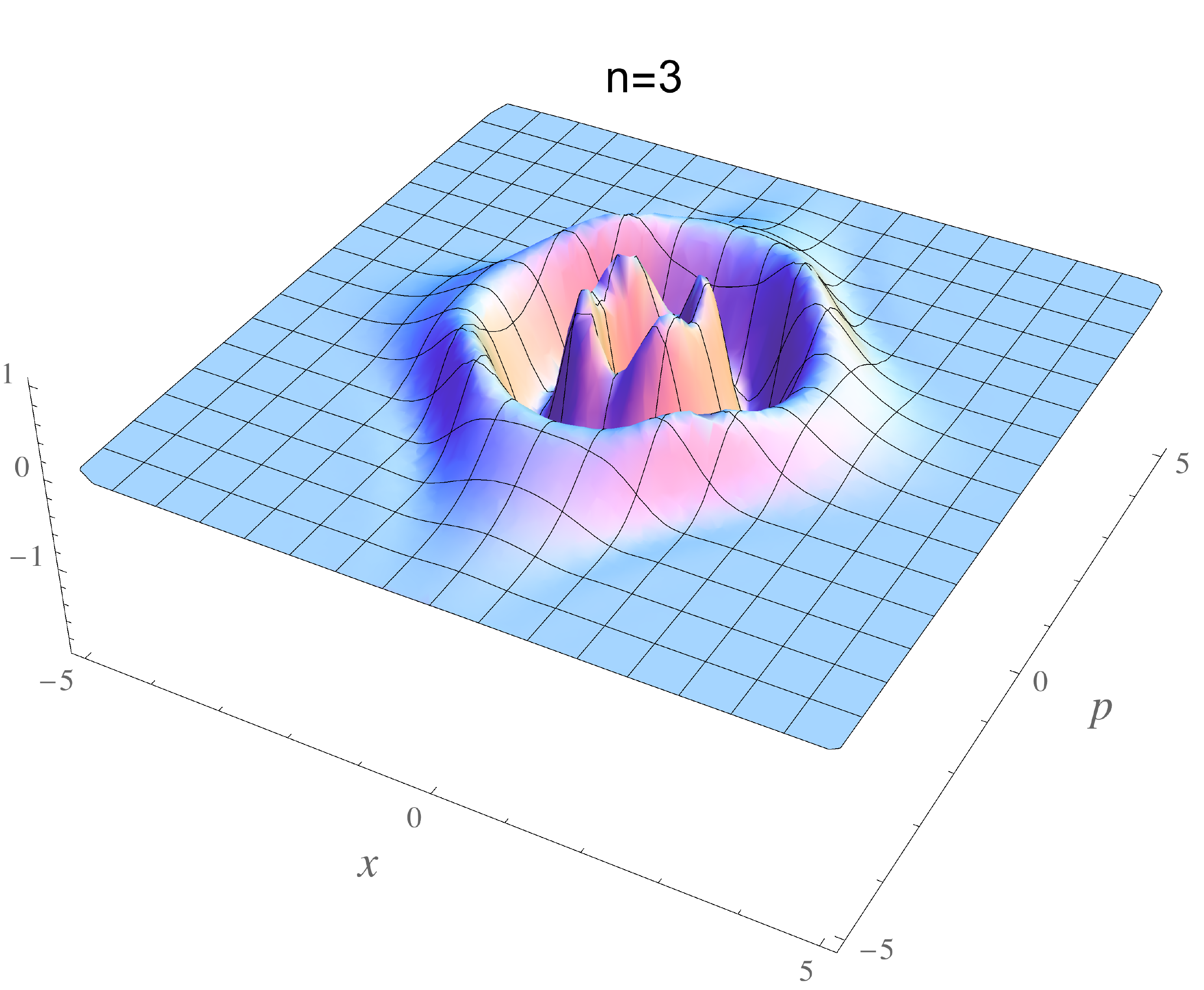}
\caption{ $n=0,1,2,3$ Wigner functions of the simple harmonic oscillator with a GUP correction for vanishing $\alpha$ (\,see equation (\ref{fpab})\,). We have set $m=\epsilon=\eta=1$ and $\beta=0.1$.  Notice that the circular symmetry is broken, but the quasi-probability distributions are unchanged by $p\to -p$, and/or $x\to -x$. }
\label{KMM WF}
\end{figure}
\end{center}

\noindent The normalization constant in (\ref{KMM solution}) is
\be
{\cal N} = \bigg(2^{\, k+m}\sqrt{\beta}\sum_{k,m=0}^{n}\lambda_{km}\tau_{km}\bigg)^{-1/2}\,,
\ee
\noindent where
\bea
\lambda_{km}=\frac{(a)_k(a)_m(-n)_k(-n)_m}{(c)_k(c)_mk!m!}
\eea
(given in terms of the Pochhammer symbol) and
\bea
\tau_{km}=\frac{i}{\mu}\big[\pi A(\mu,\kappa)-B(\mu,\kappa)\big]+\frac{i}{\kappa}\big[\pi A(\kappa,\mu)-B(\kappa,\mu)\big]\,,
\eea
with
\bea
\kappa = k-2\sqrt{q+r}+1\, ,\nnr
\mu = m-2\sqrt{q+r}+1\, ,\nnr
A(\mu,\kappa) = \frac{2^{\kappa+\mu-1}\Gamma(\mu+1)e^{i\pi\mu}\csc(\kappa+\mu)}{\Gamma(1-\kappa)\Gamma(\kappa+\mu)}\,,\nnr
B(\mu,\kappa) = 2^{\kappa+\mu-1}\,{}_2F_1(\kappa+\mu,\kappa;\kappa+1;-1).
\eea

So far, we have reviewed the results obtained by \cite{kmm}.  As a new contribution, we  {will now consider}  the Wigner functions for the wave functions just described. By numerically integrating equation (\ref{WF - p}), using equation (\ref{KMM solution}), we found the Wigner functions associated with the simple harmonic oscillator corrected by a GUP motivated by quantum gravity (Figure \ref{KMM WF}).

\begin{center}
\begin{figure}
\includegraphics[scale=0.57,angle=0]{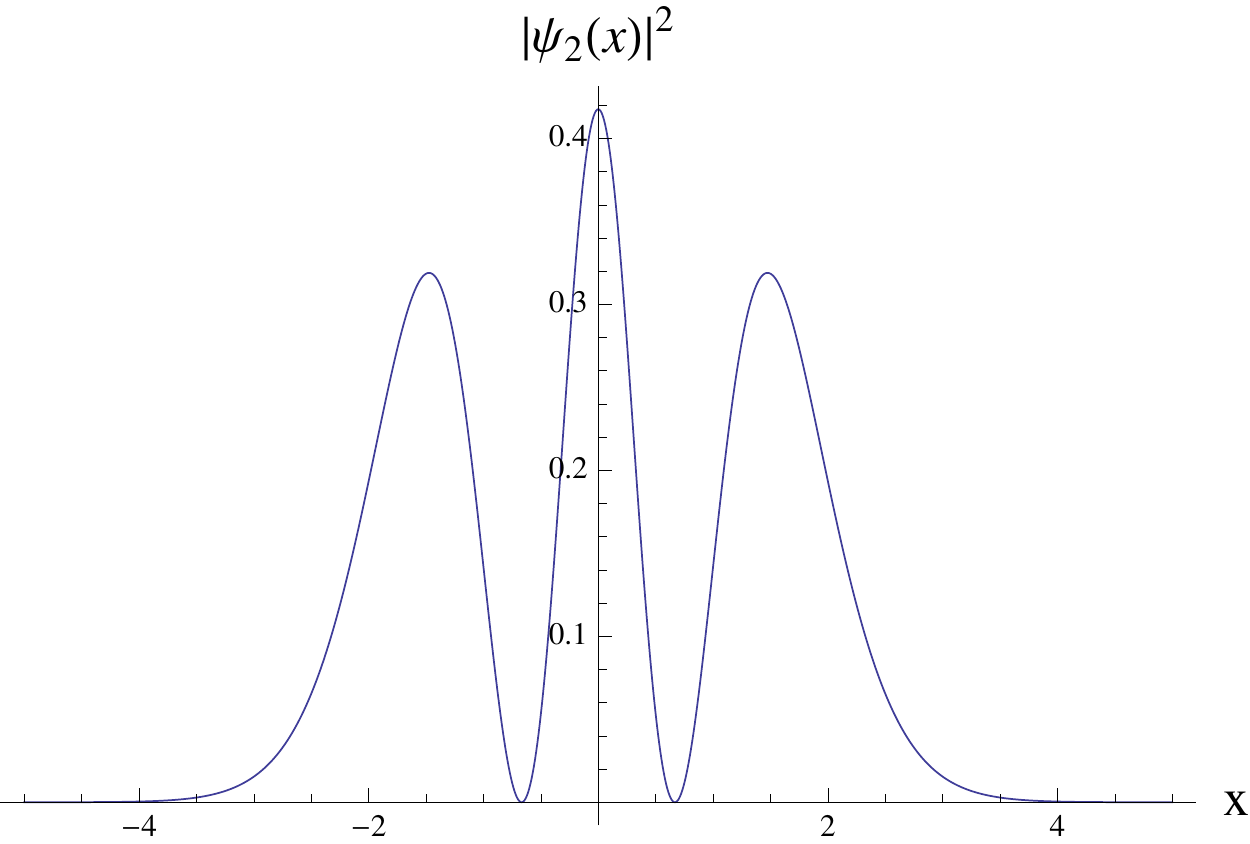}\qquad
\includegraphics[scale=0.57,angle=0]{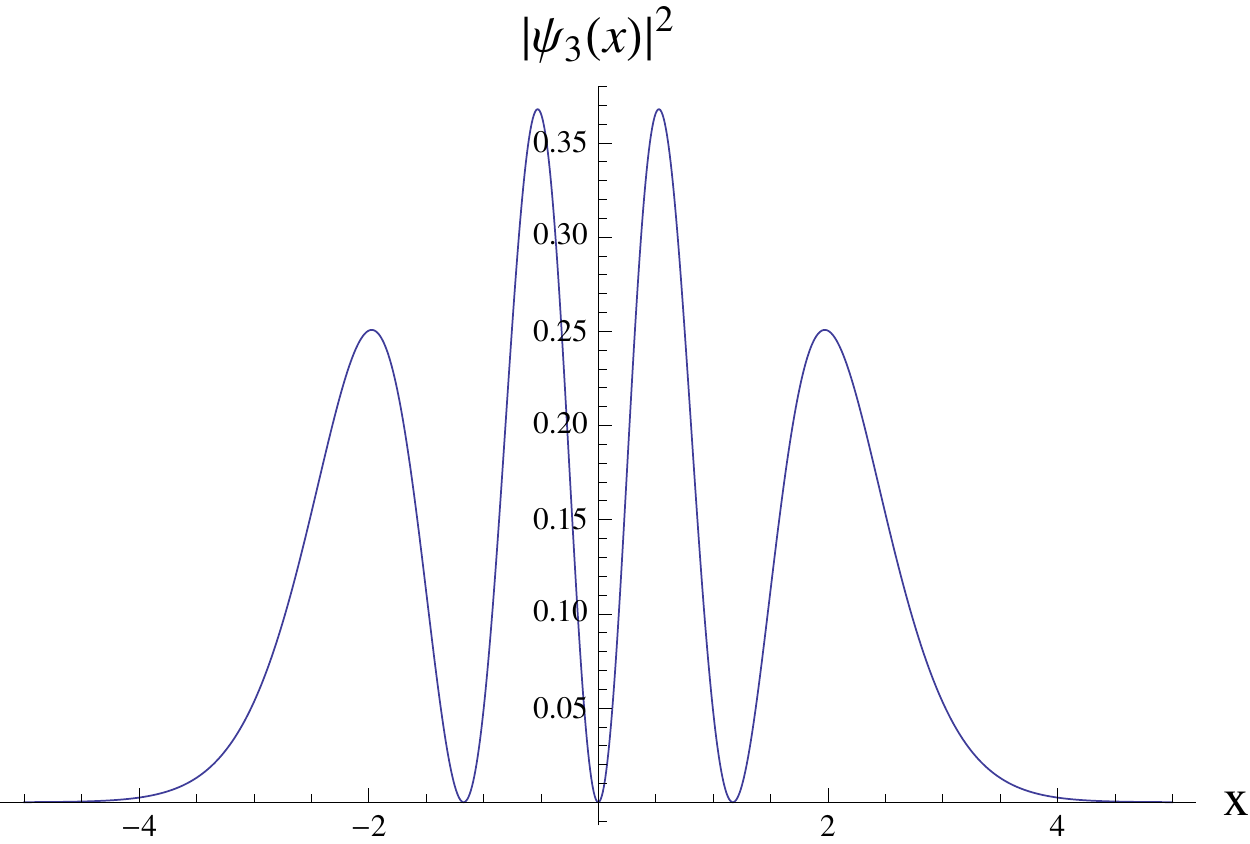}\\
\includegraphics[scale=0.57,angle=0]{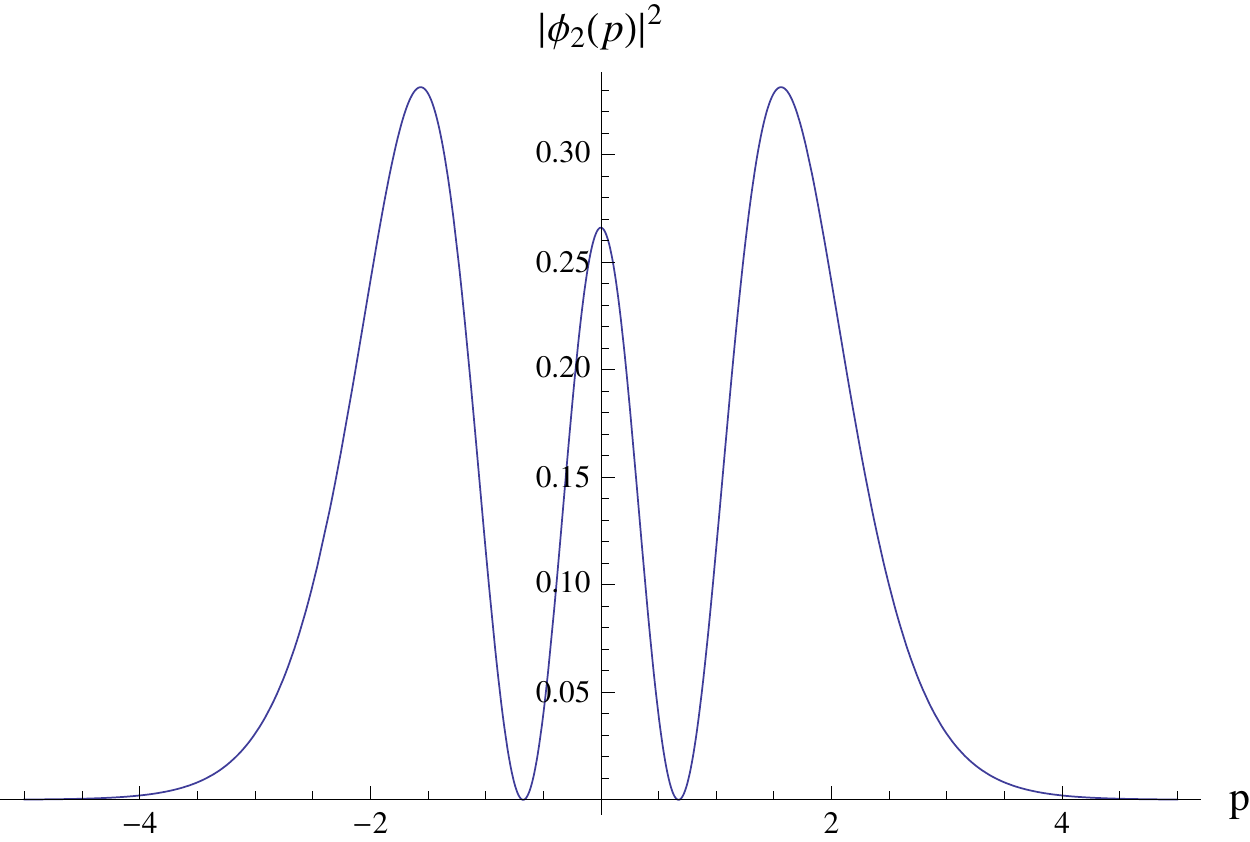}\qquad
\includegraphics[scale=0.57,angle=0]{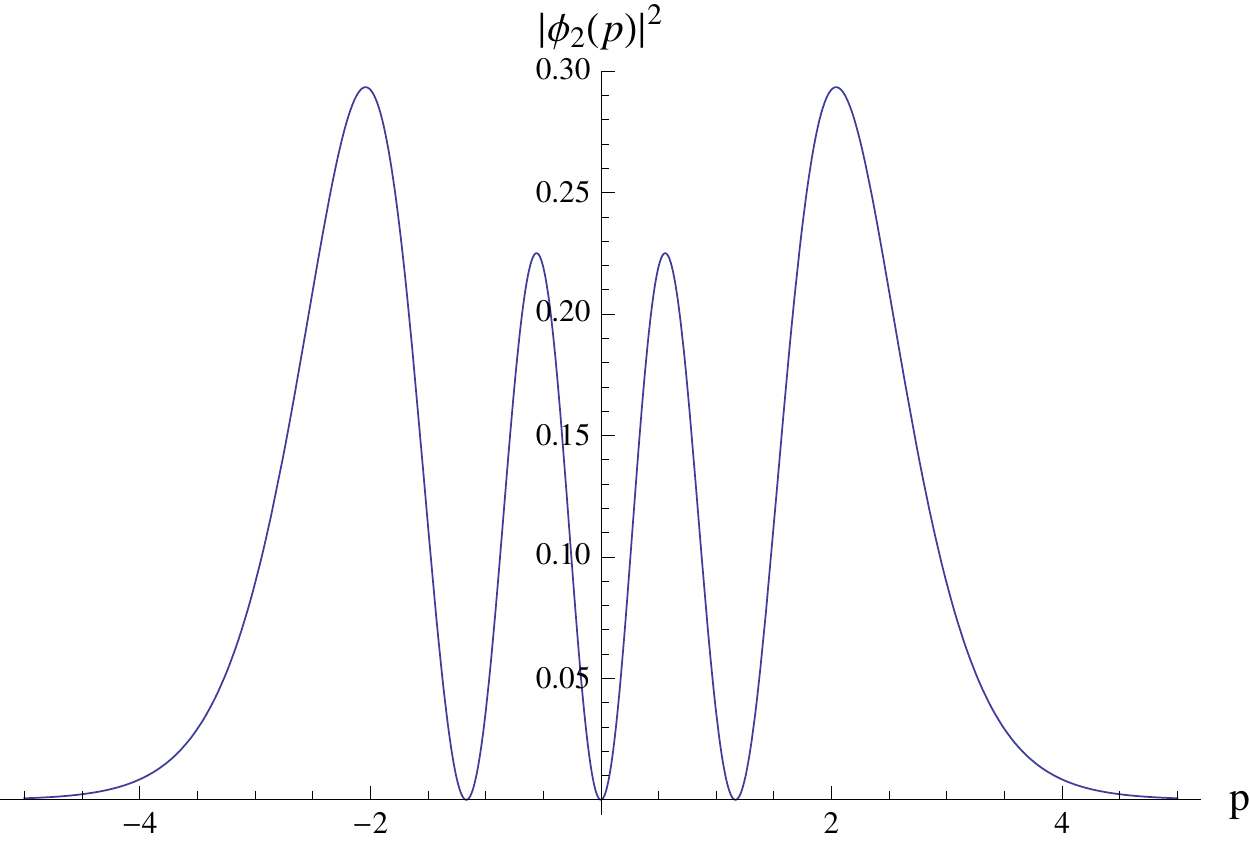}
\caption{The probability densities of the  GUP-corrected $n=2$ and $n=3$ simple harmonic oscillator energy eigenstates for $\alpha=0, \beta\not=0$. The top two plots are the $x$-space densities while the bottom plots are in $p$-space. The transformations $x\to -x$ and $p\to -p$ leave the densities invariant. We have set $m=\omega=\hbar=1$ and $\beta=0.1$.}
\label{KMM probabilities}
\end{figure}
\end{center}

Notice the deformation of the circular symmetry about the centre of the Wigner function. The quasi-probability distributions remain invariant under parity tranformations in both $x$- and $p$-space, however.  See also the probability densities plotted in  {Figure} \ref{KMM probabilities}.  {Unlike for the} regular simple harmonic oscillator, which enjoys symmetry under $x\leftrightarrow -p$, the two probability densities do not look the same.

\section{Corrections to harmonic oscillator from linear + quadratic GUP}
\label{Two-parameter GUP corrections}

 {Next, we} consider the modified Heisenberg algebra proposed in \cite{adv}, corresponding to the quantum gravity phenomenology described by (\ref{fpab}) in (\ref{GUPfp}).
%
The GUP is now
\be
\Delta x\Delta p\ \geq\ \frac{\hbar}{2}\left(1+\alpha\langle p\rangle+\beta\langle p^2\rangle\right)\ ,
\ee
and the time-independent Schr\"{o}dinger equation is
\bea
	\frac{d^2\phi(p)}{dp^2}+\frac{\alpha+2\beta p}{1+\alpha p+\beta p^2}\,\frac{d\phi(p)}{dp}+\frac{\epsilon-\eta^2p^2}{\left(1+\alpha p+\beta p^2\right)^2}\,\phi(p)\,=\,0\,,
	\label{SE - 2}
\eea
with $\epsilon$ and $\eta$ as defined above. Letting
\be \phi(p)\,=:\, \Phi(z)\, ,\quad \ z \,=\, \frac{{2\beta p+\alpha+\sqrt{\alpha^2-4\beta}}}{2{\sqrt{\alpha^2-4\beta}}}\ ,\ee we can convert equation (\ref{SE - 2}) into the form of the Riemann equation:
\be	\fl \frac{d^2\Phi(z)}{dz^2} \,+\, \frac{(2z-1)}{z\left(z-1\right)}\frac{d\Phi(z)}{dz} \,+\, \frac{q-r\left(\alpha+\sqrt{\alpha^2-4\beta}-2z\sqrt{\alpha^2-4\beta}\right)^2}{z^2\left(z-1\right)^2}\Phi(z) \,=0\ ,
\ee
where
\be
q=\frac{\epsilon}{\left(\alpha^2-4\beta\right)}\,,\hspace{1 cm}
r=\frac{\eta^2}{4\beta^2\left(\alpha^2-4\beta\right)}\ .
\ee
Solving using the Riemann $P$-symbol \cite{smirnov},
\bea
	\Phi(z)\propto\mathcal{P}\left\{\begin{array}{cccc}
		0 &  1 & \infty \\
		s & t & u_- & z\\
		-s & -t & u_+
	\end{array}\right\}
	=z^{s}(1-z)^{t}\,{}_2F_1\left(a,b;c;z\right)\ ,
\eea
where
\bea
 s = -\sqrt{-q+2r\alpha^2+2r\alpha\sqrt{\alpha^2-4\beta}-4r\beta}\ ,\nnr
t = -\sqrt{-q+2r\alpha^2-2r\alpha\sqrt{\alpha^2-4\beta}-4r\beta}\ ,\nnr
u_\pm = \frac{1}{2}\left(1\pm \sqrt{1+\frac{4\eta^2}{\beta^2}}\right)\ ,\nnr
a = u_- +s + t\,, \hspace{1 cm}b=u_+ + s + t\,,\quad
c=1+ 2s\ .
\eea

With no restrictions on $\alpha$ and $\beta$, we note that there exist non-integrable singularities.
However,  {if we assume} $\alpha^2<4\beta$, we find $s=t^*$,  {thus, eliminating this problem}.

To analyze the asymptotics of the wave function, we use
$z^vw^{v^*}=z^{x+i y}w^{x-i y}=(zw)^xe^{i y\ln\left(\frac{z}{w}\right)},$ valid for arbitrary $z,v,w\in\C,$ $x,y\in\R$.
We find
\bea
\Phi(z)\propto\left[z(1-z)\right]^{\Re(s)}
e^{i\,\Im(s)\ln\frac{z}{1-z}}\, {}_2F_1\left(a,b;c;z\right)\ .
\eea
Since we want to ensure that the square of the norm of the wave function converges when integrated, we consider two cases:   1) $a=-n$ and 2) $b=-n$;  here $n\in\Z^+\cup\{0\}$ so that the Gauss hypergeometric function reduces to a polynomial of order $n$. For $a=-n$, we find:
\bea
b = -n+\sqrt{1+\frac{4\eta^2}{\beta^2}}\,,\nn\\
c = 1-n-u_-+2i\,{\Im}(s),\nn\\
{\Re}(s) = -\frac{n+u_-}{2}\,,
\eea
so that
\bea
\fl \Phi_1(z)\ =\ \left[z(1-z)\right]^{-\frac{n+u_-}{2}}e^{i\,{\Im}(s)\,
\ln\frac{z}{1-z}}\, {}_2F_1\left(-n,-n+\sqrt{1+\frac{4\eta^2}{\beta^2}};1-n-u_-+2i\,{\Im}(s);z\right)\nn\\
 \sim z^{-u_-}\  ,\   \   {\rm for\  large}\  \vert z\vert\  \  .
\eea
For $b=-n$:
\bea
a = -n-\sqrt{1+\frac{4\eta^2}{\beta^2}}\,,\nn \\
c = 1-n-u_++2i\,{\Im}(s)\,,\nn \\
{\Re}(s) = -\frac{n+u_+}{2}\,,
\eea
so that
\bea
\fl\Phi_2(z)\ =\ \left[z(1-z)\right]^{-\frac{n+u_+}{2}}\,
e^{i\,{\Im}(s)\ln\frac{z}{1-z}}\, {}_2F_1\left(-n-\sqrt{1+\frac{4\eta^2}{\beta^2}},-n;1-n-u_+ + 2i\,{\Im}(s);z\right)\nn\\
 \sim z^{-u_+}\  ,\  \  {\rm for\ large}\  \vert z\vert\ \  .
\eea
 We see that, as $z\to\infty$, $\Phi_1$ diverges, thus $\phi(p) \propto \Phi_2(z)$,
\bea
\fl \phi(p)\, \propto\,
\left[\alpha^2-4\beta-(2p\beta+\alpha)^2\right]^{-\frac{n+u_+}{2}}
\exp\left\{i\,{\Im}(s)\,\ln\left[\frac {2\,p\beta+\alpha+\sqrt {{\alpha}^{2}-4\,\beta}}{\sqrt {{\alpha}^{2}-4\,\beta}-2\,p\beta-\alpha}\right]\right\}\nn\\
\fl {}\quad \times\, {}_2F_1\left(-n-\sqrt{1+\frac{4\eta^2}{\beta^2}},-n;1-n-u_++2\,{\Im}(s); \, \frac{2\beta p+\alpha+\sqrt{\alpha^2-4\beta}}{2\sqrt{\alpha^2-4\beta}}\right)\  .
\label{DRWwf}\eea
%
%
\begin{center}
\begin{figure}[t!]
\includegraphics[scale=1.3, angle=0]{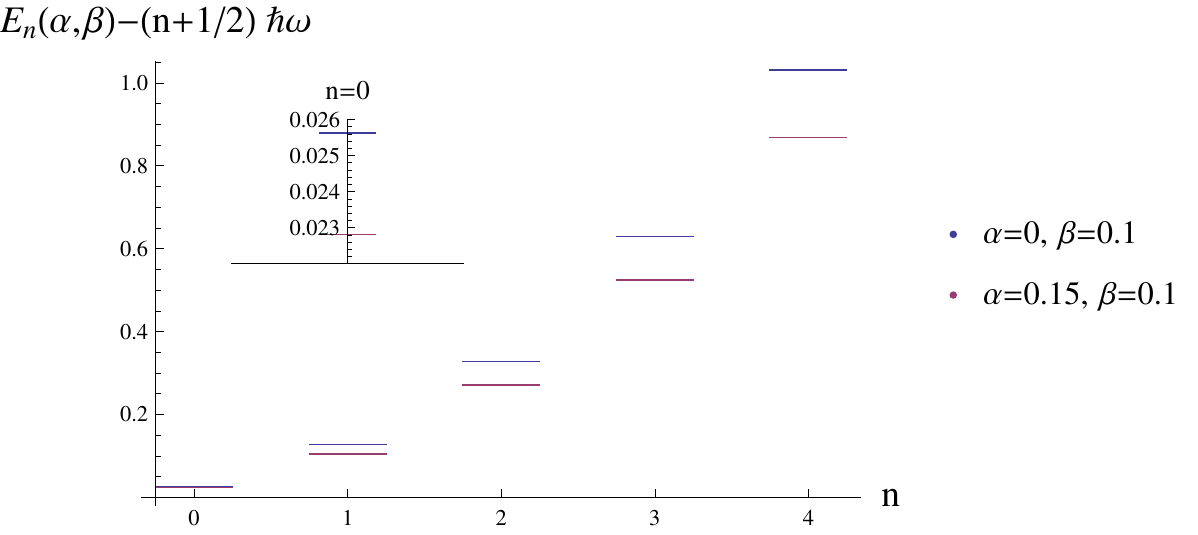}
\caption{Energy levels are indicated for the simple harmonic oscillator with the 1-parameter GUP correction (blue, $\alpha=0$, $\beta=0.1$), and with the 2-parameter GUP correction (red, $\alpha=0.15$, $\beta=0.1$).  The differences between the corrected and uncorrected energies are shown; in all cases they are larger for the 1-parameter correction. The inset shows the small energy difference between the 2 cases for $n=0$.  We have set $m=\epsilon=\eta=\hbar=1$.}
\label{EnergyDiagram}
\end{figure}
\end{center}

\subsection{GUP Corrected Energy Spectrum}
Using
\be
\epsilon = \frac{2E}{\hbar^2\omega^2}\ ,\qquad
q = \frac{\epsilon}{(\alpha^2-4\beta)}\ ,\qquad
-n = b = u_++s+t\ ,
\ee
\noindent we find that
\bea
\fl E_n&\ =\ \frac{\alpha^2m\omega^2\hbar^2\left[n\xi\left(2\sigma\left\{1+\theta+n[3\xi+2\theta]\right\} - n\xi\sigma^2\left\{1+\theta+2n[\xi+\theta]\right\} - 8\right)-2\right]}{16\left(\sigma n^2+\sigma n -1\right)^2}\nn\\
\fl &\qquad\qquad\qquad +\ \frac{4\beta m\omega^2\hbar^2\left[1+\theta+2n(\xi+\theta)\right] \left[n\xi\sigma-1\right]^2}{16(\sigma n^2+\sigma n -1)^2}\ ,
\label{EnAB}\eea
where $\displaystyle{\theta=\sqrt{1+\frac{4}{\sigma}}}$, $\xi=1+n$, and $\sigma=(m\beta\omega\hbar)^2$. Note that for $\alpha\to0$, the spectrum (\ref{EnB}) is recovered.
Subsequently taking $\beta\to0$, the energies reduce to the expected $E_n=\left(n+1/2\right)\hbar\omega$.
\begin{center}
\begin{figure}
\includegraphics[scale=0.26,angle=0]{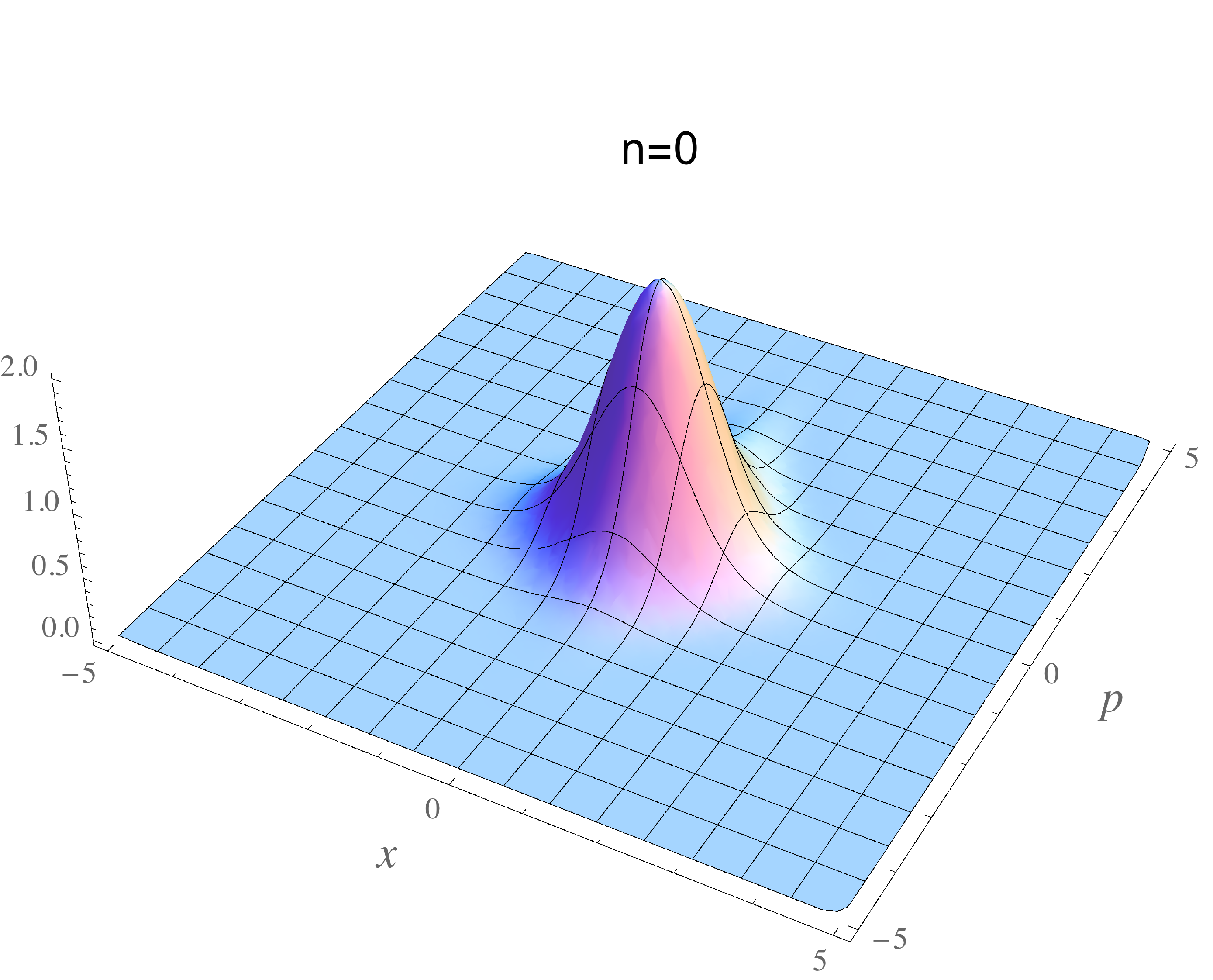}\qquad
\includegraphics[scale=0.26,angle=0]{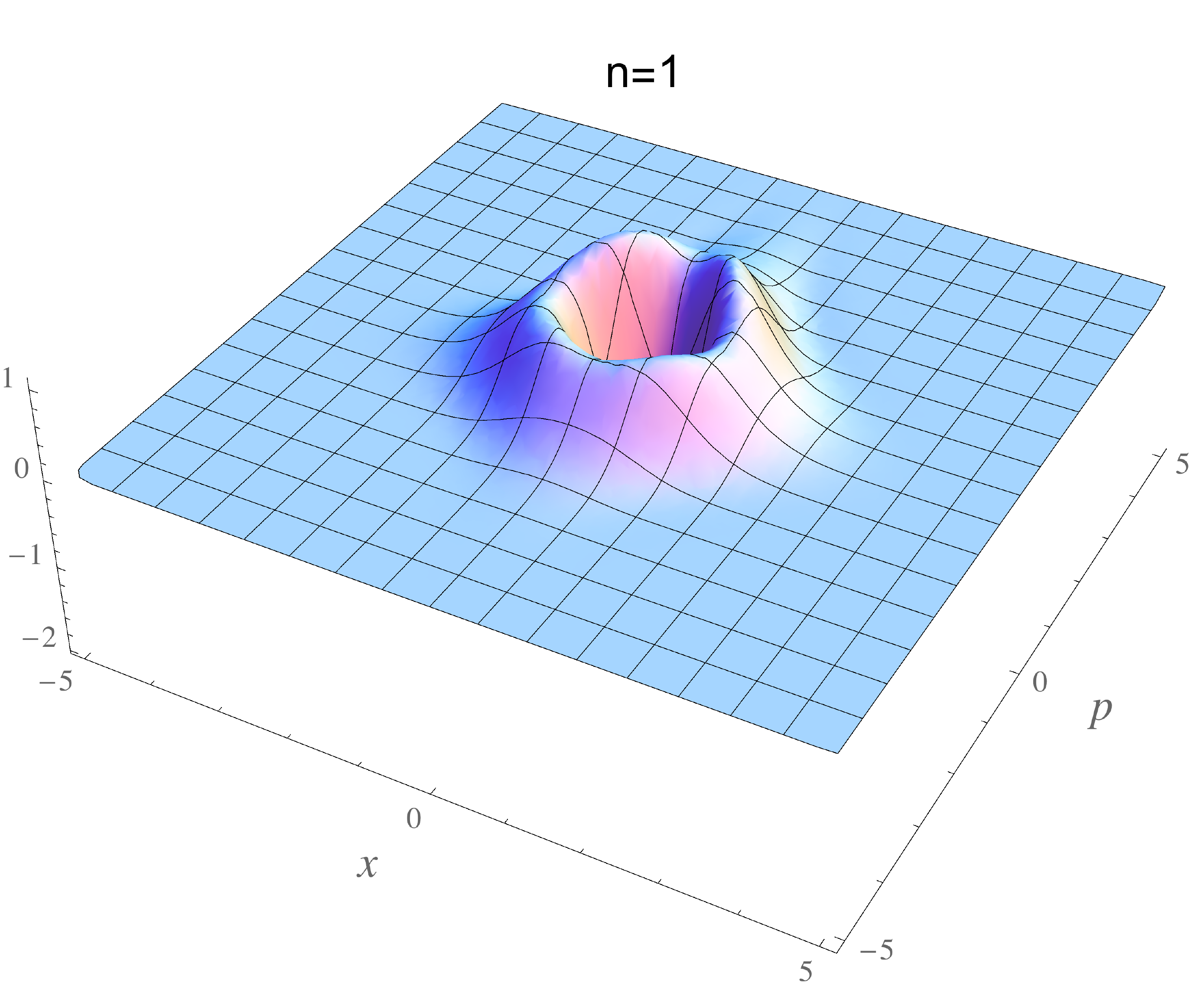}\\
\includegraphics[scale=0.26,angle=0]{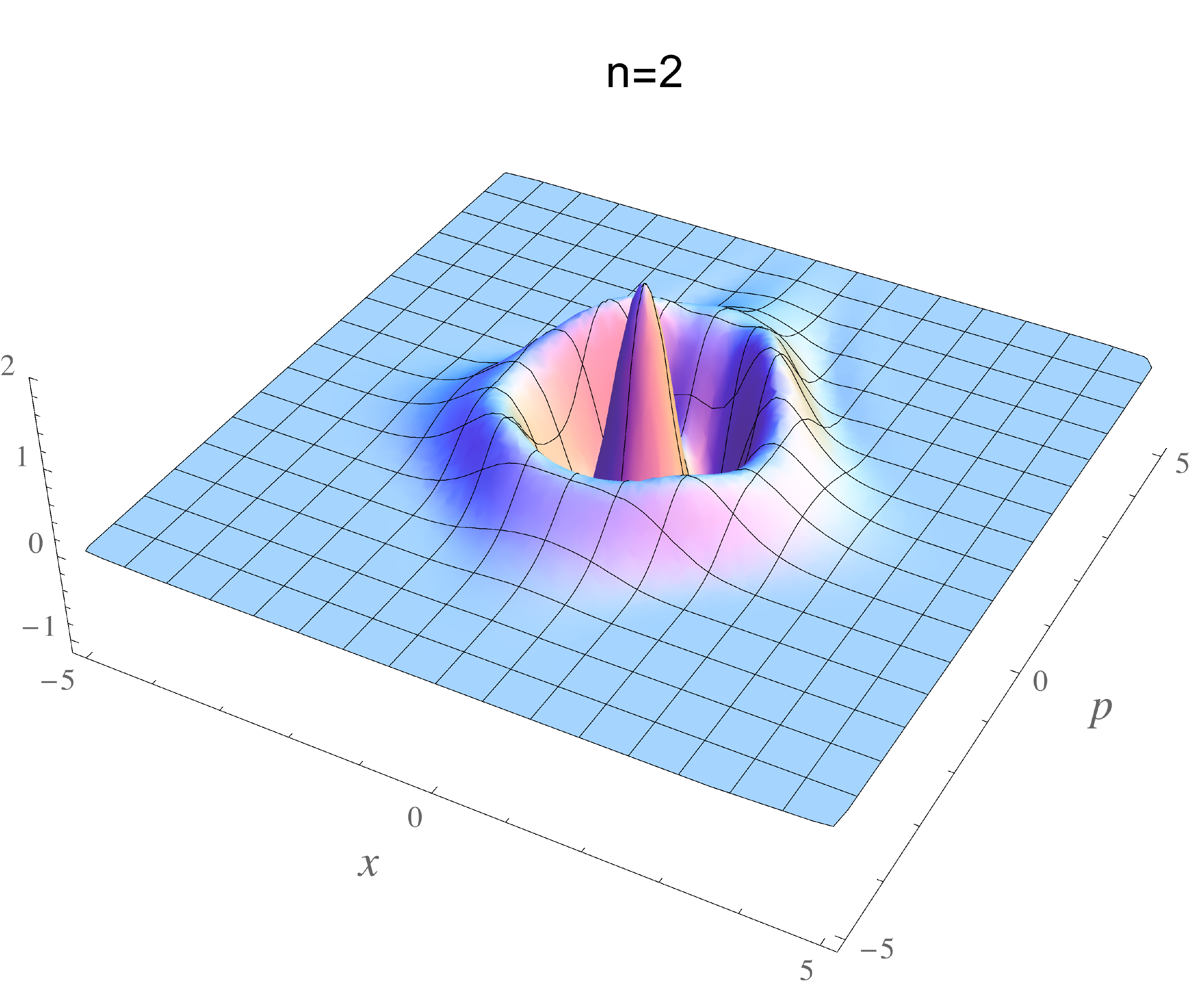}\qquad
\includegraphics[scale=0.26,angle=0]{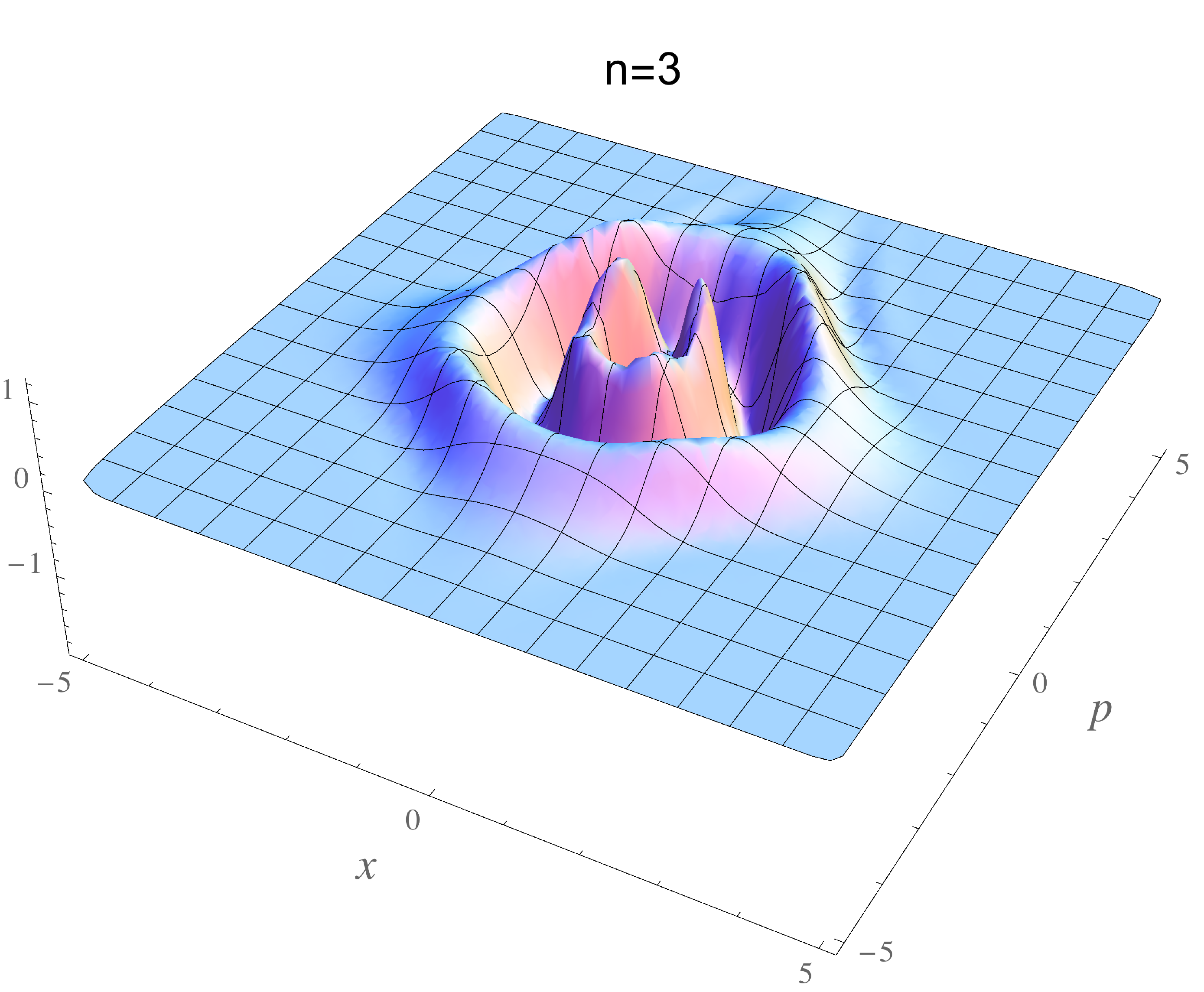}
\caption{GUP-corrected Wigner functions with $\alpha=0.15$ and $\beta=0.1$. We have set $m=\epsilon=\eta=\hbar=1$.}
\label{generalized QGWF}
\end{figure}
\end{center}

\begin{center}
\begin{figure}
\includegraphics[scale=0.565,angle=0]{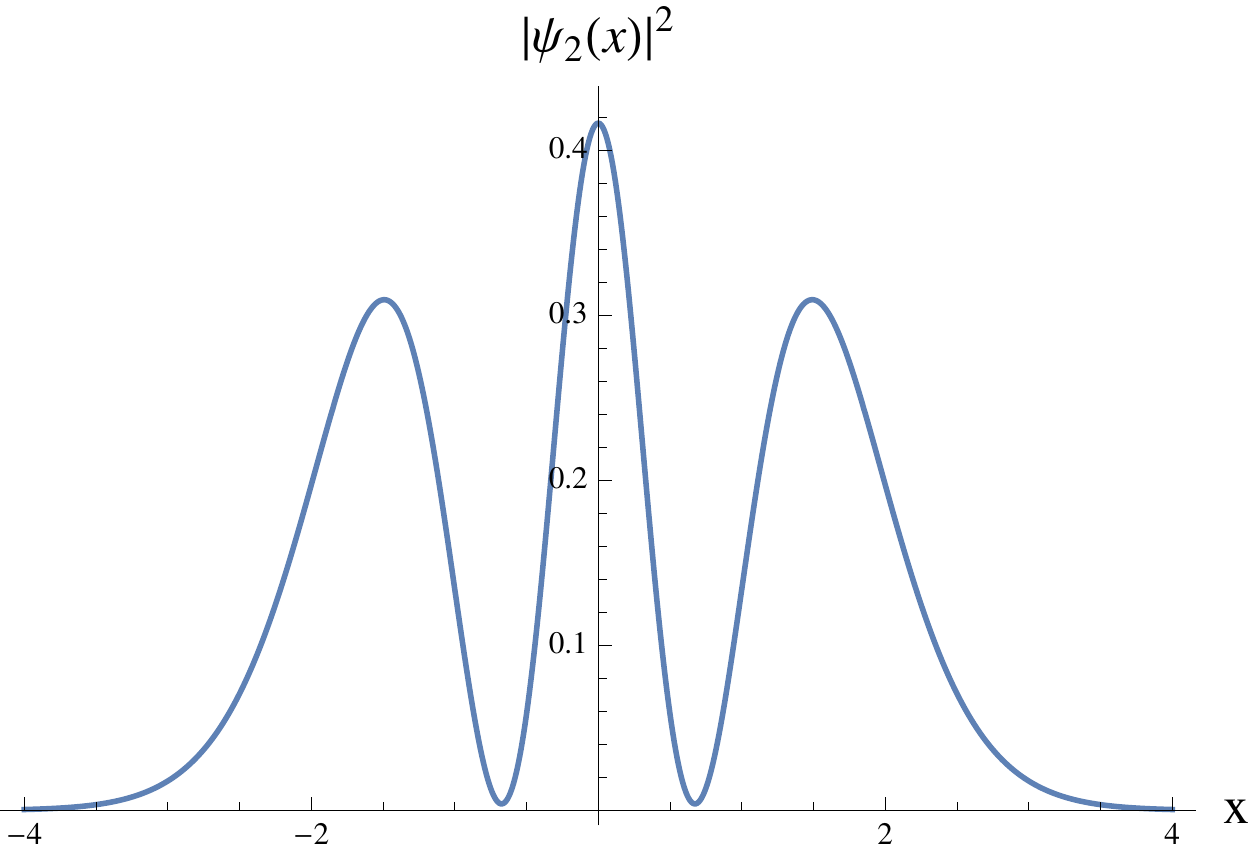}\qquad\quad
\includegraphics[scale=0.565,angle=0]{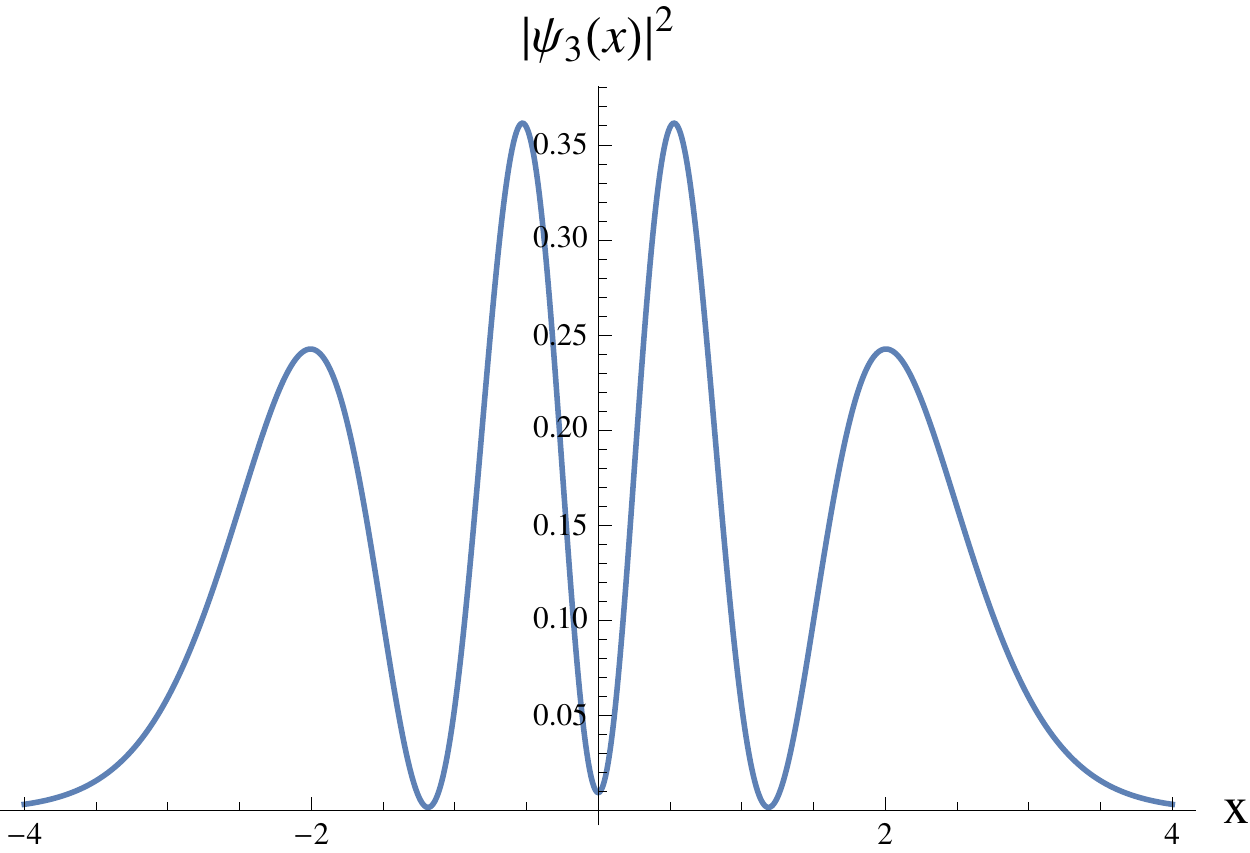}\\
\includegraphics[scale=0.565,angle=0]{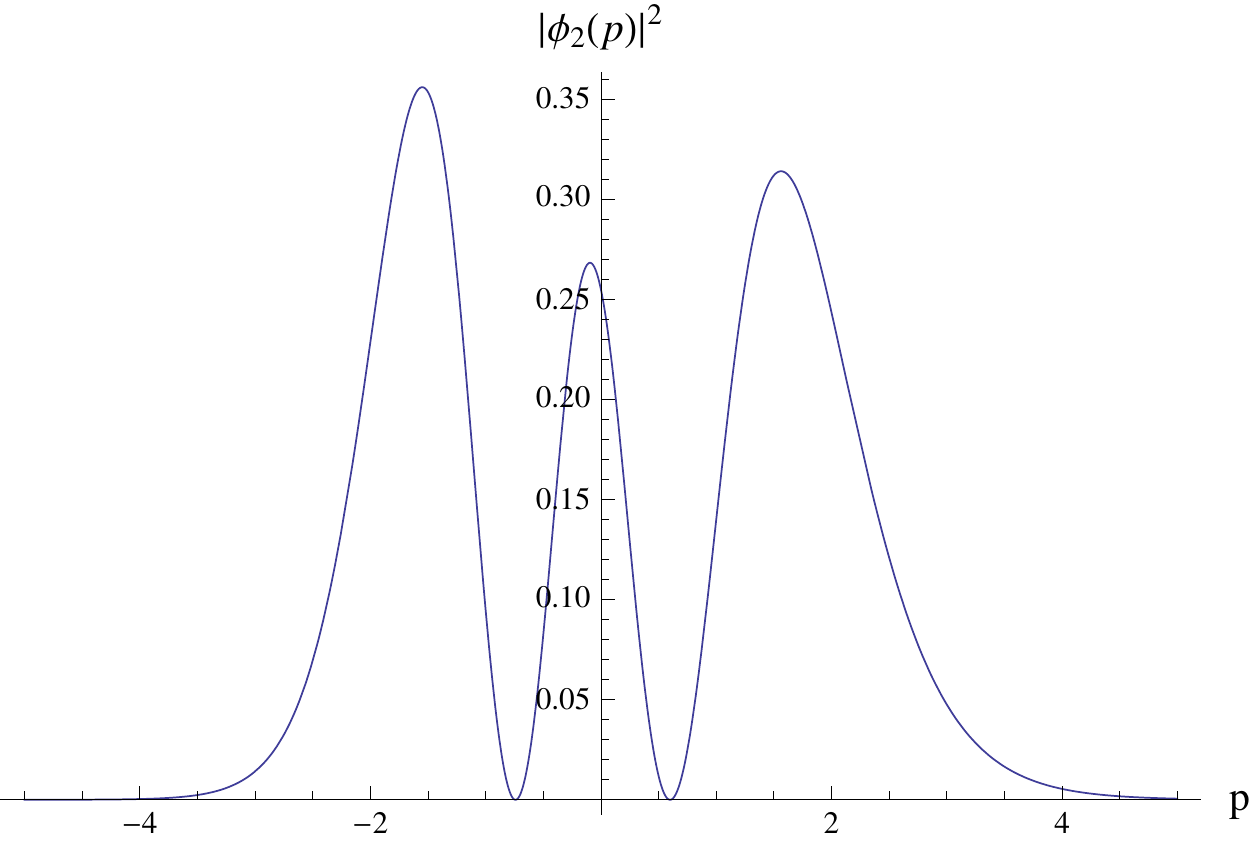}\qquad\quad
\includegraphics[scale=0.565,angle=0]{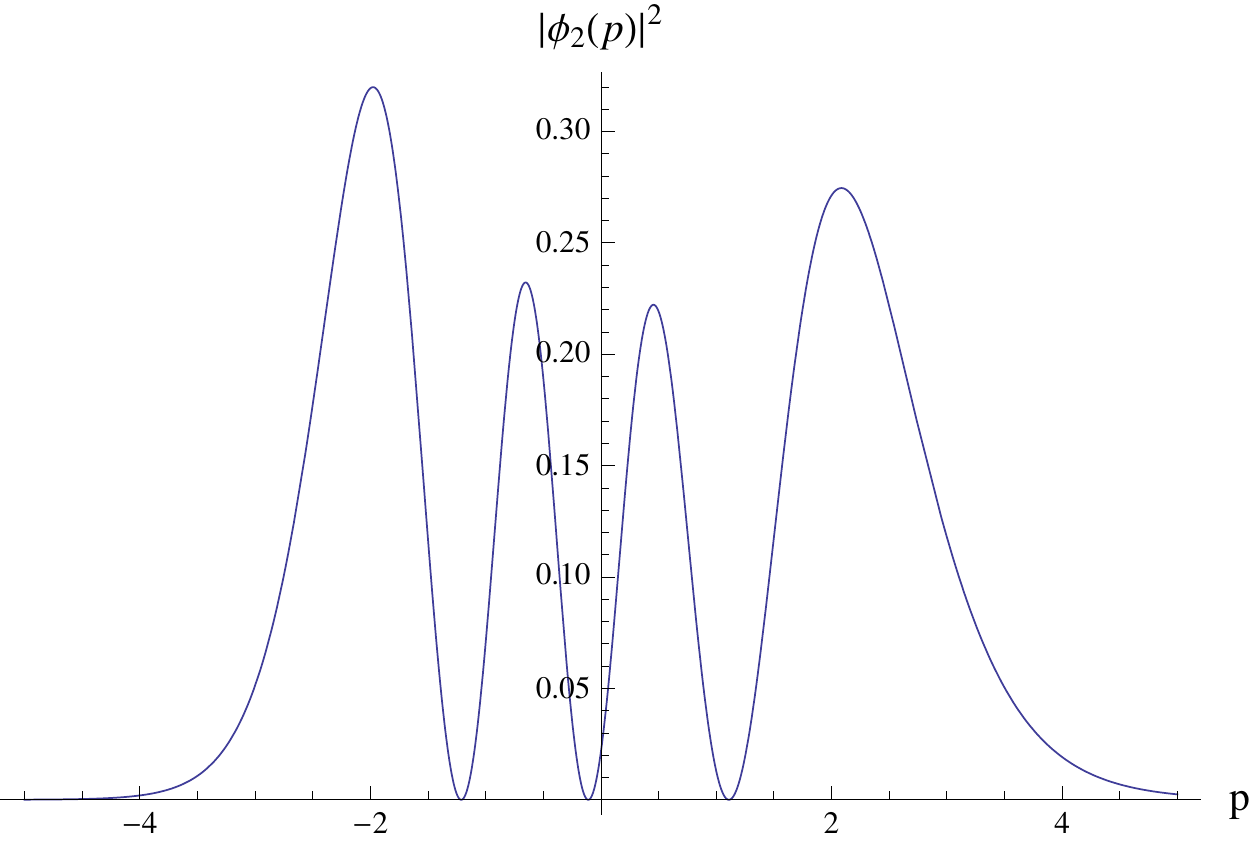}
\caption{GUP-corrected probability densities with $\alpha=0.15$ and $\beta=0.1$ for $n=2$ and $n=3$. The top (bottom) 2 plots show $x$-space ($p$-space). The symmetry $x\to -x$ ($p\to -p$) is intact (broken). We have set $m=\epsilon=\eta=\hbar=1$.}
\label{generalized probabilities}
\end{figure}
\end{center}
Figure \ref{EnergyDiagram} depicts the lowest energies ($n=0,\ldots , 4$) of the spectra for $\alpha=\beta=0$ (unperturbed simple harmonic oscillator), $\alpha=0$, $\beta=0.1$ (simple harmonic oscillator with 1-parameter GUP correction), and $\alpha=0.15$, $\beta=0.1$ (simple harmonic oscillator with 2-parameter GUP correction). Notice that, even for small $n$, the difference in energy levels grows rapidly. Also, while the perturbations raise the energies in both cases, the effect is smaller when both $\alpha,\beta\not=0$.

\subsection{GUP Corrected Wigner Functions}
The quantum gravity-modified Wigner functions for a GUP (\ref{fpab}) with both $\alpha$ and $\beta$ non-vanishing (Figure \ref{generalized QGWF}) exhibit a modified deformation from that for $\alpha=0$ (Figure \ref{KMM WF}), with the difference becoming clearer  {as $n$ becomes larger}. While invariance under $x\to -x$ remains intact, symmetry under $p\to -p$ is broken.

Correspondingly, the probability densities for the 2-parameter GUP correction differ from those for the 1-parameter case (Figure \ref{generalized probabilities}). Note that the disappearance of the symmetry between the $x$- and $p$-space probability densities is more pronounced. Further, though the $x$-space probability densities are symmetric about $x=0$, there is a greater probability of finding a particle in the region $p>0$. This is consistent with the broken $p$-parity.\vskip1.6cm

\section{Conclusion}
\label{Outlook}
We first point out our main results.  For the GUP specified by $[\hat x, \hat p] = i\hbar(1 + \alpha\hat p +\beta\hat p^2)$, we have derived the wave functions  (\ref{DRWwf}) for the simple harmonic oscillator in momentum space, and energy spectrum (\ref{EnAB}).  These generalize the results  (\ref{KMM solution}) and (\ref{EnB}) of \cite{kmm}, to $\alpha\not=0$ .

The wave functions, both old and new, allowed us to investigate for the first  {time,} the corresponding Wigner functions  {in} phase space, by implementing (\ref{WF - p}) numerically. We have included several plots of the Wigner functions, that illustrate the effects of the GUP corrections, both when $\alpha$ is zero, and non-zero.  Significant changes to the uncorrected Wigner functions (see Figure \ref{SHO WF}) are found, that  {intensify} with increasing oscillator energy, and break the circular symmetry (dependence on only $x^2+p^2$) in phase space (see Figures \ref{KMM WF} and \ref{generalized QGWF}).  The probability densities in both coordinate and momentum space are also illustrated in Figures. \ref{KMM probabilities} and \ref{generalized probabilities}.  For $\alpha=0$, $\beta\not=0$, invariance under both $x\to-x$ and $p\to-p$ remain.  For both $\alpha,\beta\not=0$, only the parity symmetry $x\to -x$ survives.

Our  {supposition} is that these, or similar corrections to Wigner functions may be observable.  The Wigner functions corresponding to quadratures of electromagnetic fields can be reconstructed in quantum optical systems, either by homodyne detection in cavities and then by a Radon inverse transform \cite{opticsI}, or directly via photon-number-resolving detection \cite{opticsII}.  It may therefore be possible to measure quantum gravity corrections to the Wigner function in similar systems.  {Interestingly, the techniques that may be useful are also pertinent to the study of the classical limit in quantum mechanics} \cite{opticsI}. We hope to study  {this} in detail and report elsewhere.

\vs{.6cm}

\ack
This work was supported by Discovery Grants (SD, MAW) and an Undergraduate Student Research Award (MPGR) from the Natural Sciences and Engineering
Research Council of Canada. MPGR was also supported by the George Ellis Research Scholarship from the University of Lethbridge.
%


\end{document}